\documentclass[a4paper,12pt,nofootinbib]{article}
\usepackage[affil-it]{authblk}
\usepackage{amsmath,amssymb,amsfonts}
\usepackage{graphicx,epsfig, subfigure}
\usepackage{color}
\usepackage{slashed}
\usepackage{epigraph}
\usepackage{pifont}
\usepackage{slashed}
\usepackage{latexsym}
\usepackage{listings}
\usepackage{dcolumn}
\usepackage{comment,latexsym} 
\usepackage{verbatim}
\usepackage{longtable,tabularx}
\usepackage{dcolumn}
\usepackage{chngpage}
\usepackage{hyperref}
\usepackage{cite}

\numberwithin{equation}{section}
\def\bea{\begin{eqnarray}}
\def\eea{\end{eqnarray}}

\newcommand{\bear}{\begin{array}}
\newcommand{\ear}{\end{array}}
\newcommand{\nn}{\nonumber}

\def\OMIT#1{{}}
\newcommand{\lsim}{\mathrel{\rlap{\lower4pt\hbox{\hskip1pt$\sim$}}
    \raise1pt\hbox{$<$}}}       
\newcommand{\gsim}{\mathrel{\rlap{\lower4pt\hbox{\hskip1pt$\sim$}}
    \raise1pt\hbox{$>$}}}

\newcommand{\Tr}{{\rm Tr}}
\newcommand{\be}{\begin{eqnarray}}
\newcommand{\ee}{\end{eqnarray}}
\newcommand{\ba}{\begin{eqnarray}}
\newcommand{\ea}{\end{eqnarray}}

\newcommand{\LL}{\mathcal{L}}

\newcommand{\OO}{\mathcal{O}}

\title{
Selfish Dark Matter \\
\small Spontaneous baryogenesis from extremely light asymmetric dark matter}
\author{Raffaele Tito D'Agnolo, Anson Hook}
\affil{School of Natural Sciences, Institute for Advanced Study, Princeton, NJ, 08540}
\date{}

\begin{document}

\maketitle

\begin{abstract}
We present a mechanism where a particle asymmetry in one sector is used to generate an asymmetry in another sector. The two sectors are not coupled through particle number violating interactions and are not required to be in thermal contact with each other.  When this mechanism is applied to baryogenesis in asymmetric dark matter models, we find that the dark matter particles can be extremely light, e.g. much lighter than an eV, and that in some cases there is no need to annihilate away the symmetric component of dark matter. We discuss a concrete realization of the mechanism with signals in direct detection, at the LHC, at $B$-factories or future beam dump experiments.
\end{abstract}

\newpage

\section{Introduction}

Since the discovery of dark matter (DM), there have been two paradigms developed to explain its current number density.  The first is the idea of thermal freeze-out. When the dark matter annihilation rate falls below the Hubble expansion rate, a relic abundance results.  The second paradigm postulates the existence of an asymmetry between the number densities of dark matter particles and anti-particles.  It is then impossible for the dark matter to be completely annihilated away. In the first case, the dark matter and baryon energy densities are completely unrelated and their approximate equality today ($\rho_{DM}/\rho_B \approx 5$~\cite{Ade:2013ktc}) is just an accident. On the other hand, asymmetric dark matter models can naturally relate the two quantities by either generating an asymmetry in one sector and then transferring it to the other sector, or by generating both at the same time.  
Cases in which the asymmetry is transferred via electroweak sphalerons~\cite{Barr:1990ca, Barr:1991qn, Kaplan:1991ah, Gudnason:2006ug, Gudnason:2006yj, Ryttov:2008xe, Foadi:2008qv, Frandsen:2009mi, Frandsen:2011kt, Lewis:2011zb} or other baryon number violating interactions~\cite{Kaplan:2009ag, Cui:2011qe, Dulaney:2010dj, Perez:2013nra, Dodelson:1993je, Hooper:2004dc, Falkowski:2009yz,Cohen:2010kn, Haba:2011uz, Kang:2011wb, Okada:2012rm, Choi:2013fva, Cheung:2011if} have been explored extensively in the literature.  
These models typically predict $\Omega_{DM}/\Omega_B \approx m_{DM}/m_B$, once the symmetric component of dark matter is annihilated away. Therefore accounting for the approximate coincidence in energy densities requires annihilating away efficiently the symmetric component of dark matter and tying the dark matter mass to the baryon mass scale\footnote{Exceptions include models where the dark matter becomes non-relativistic when electroweak sphalerons are still in thermal equilibrium~\cite{Barr:1990ca, Barr:1991qn} and where CP, baryon and dark matter number violation is present in both sectors, generating the two asymmetries simultaneously~\cite{Falkowski:2011xh}.}. 

In this paper, we present a new mechanism where an asymmetry is used to generate another asymmetry. When applied to asymmetric dark matter models, it can produce a large hierarchy between the dark matter and proton masses and in some cases, removes the need to annihilate away the symmetric component of dark matter. If the asymmetry is first generated in the dark sector and then used to create a baryon number asymmetry, we show that it is possible to accommodate sub eV asymmetric dark matter candidates, a possibility not present in existing models in the literature.  

Rather than sharing an asymmetry between two sectors, we consider using the asymmetry in the dark matter sector to spontaneously \emph{generate} a new asymmetry in the visible sector\footnote{A similar mechanism was discussed in Ref.~\cite{Blum:2012nf, Servant:2013uwa} where the interactions of the dark matter sector with the Higgs boson is such that in thermal equilibrium a non-zero B-L charge induces a non-zero value for the non-conserved dark matter charge.}. We call this new scenario ``selfish" dark matter. In more detail, first an asymmetry in the dark matter sector is produced via any of the known mechanisms: decays of heavy particles~\cite{Nanopoulos:1979gx}, Affleck-Dine flat directions~\cite{Affleck:1984fy}, spontaneous darkogenesis~\cite{MarchRussell:2011fi, Kamada:2012ht}, dark electroweak baryogenesis~\cite{Dutta:2010va, Shelton:2010ta, Petraki:2011mv, Walker:2012ka} or black holes~\cite{Hook:2014mla}. Next, we assume that the Lagrangian of the theory contains the operator 
\be
\frac{J^\mu_{DM}\left(g_- J_{\mu}^{B-L}+ g_+ J_{\mu}^{B+L}\right)}{\Lambda^2}~\label{eq:JJ}\, .
\ee
These interactions do not violate baryon or dark matter number.  Instead they source a chemical potential for baryons proportional to the asymmetric dark matter number density $\mu \sim \langle J^0_{DM} \rangle/\Lambda^2 = n_{DM}/\Lambda^2$.  In the presence of a chemical potential, thermal equilibrium carries a net charge
\bea
n_b - n_{\bar b} \sim \mu T^2 \sim \frac{n_{DM} T^2}{\Lambda^2}\, .
\eea
This alone is not enough to generate a baryon asymmetry. However if explicit baryon or lepton number violation, such as electroweak sphalerons or right handed neutrino masses, are in equilibrium, then an asymmetry in the dark matter sector generates an asymmetry in the visible sector proportional to the temperature at which these interactions freeze-out $n_b - n_{\bar b} \sim n_{DM} T_{B\pm L}^2/\Lambda^2$.  While the source of our chemical potential is different, this is the approach of spontaneous baryogenesis~\cite{Cohen:1987vi, Cohen:1988kt}.  While the focus was different, Ref.~\cite{Banks:2006xr} also used this operator to generate spontaneous baryogenesis.

This simple mechanism has three important features. First, since $T_{B\pm L}$ and $\Lambda$ are a priori unrelated, it is possible to generate the correct ratio of the DM and baryon energy densities for dark matter masses far from the GeV scale. Second, the asymmetry in the visible sector is generated by the background value of an operator, so thermal contact between the Standard Model (SM) and the dark sector is not required for successful baryogenesis. This opens the possibility of evading Big Bang Nucleosynthesis and structure formation bounds on the dark matter mass and having extremely light asymmetric dark matter.  Finally, in realizations of the mechanism where the dark and visible sector are not in thermal contact, we do not need to annihilate away the symmetric component of dark matter.   We can use mechanisms such as Affleck-Dine baryogenesis in the dark sector, which naturally produce large asymmetric components, without annihilating away the remaining symmetric component.

In the Standard Model, electroweak sphalerons break $B+L$ so we can either consider theories with only $B+L$ violation or theories that break both $B+L$ and $B-L$. In the first case, baryon number violation decouples when the electroweak sphalerons freeze-out at a temperature $T_{S}\approx 10^2$~GeV.  
This realization of the mechanism features an ``Asymmetric WIMP miracle".  If the dark matter mass is around a GeV, the interactions in Eq.~(\ref{eq:JJ}) are suppressed by the TeV scale and can give signatures at direct detection experiments, $B$-factories, the LHC and beam dump experiments. In the second case, the highest decoupling temperature ($T_{B-L}$) dominates the generation of the asymmetry. 

In Section~\ref{Sec: Toy}, we present a simple toy example which has all of the features of our model.  In Section~\ref{sec:chem}, we show how the dark matter asymmetry generates a chemical potential and find the region of parameter space of the model compatible with the observed baryon number abundance.
In Section~\ref{sec:constraint}, we discuss in more detail why our mechanism does not require thermal contact between the two sectors and how extremely light dark matter can be accommodated.  In Section~\ref{sec:models}, we discuss realizations of the mechanism with interesting phenomenological features.

\section{A Toy Model} \label{Sec: Toy}
\begin{figure}[!t]
\begin{center}
\includegraphics[width=0.5\textwidth]{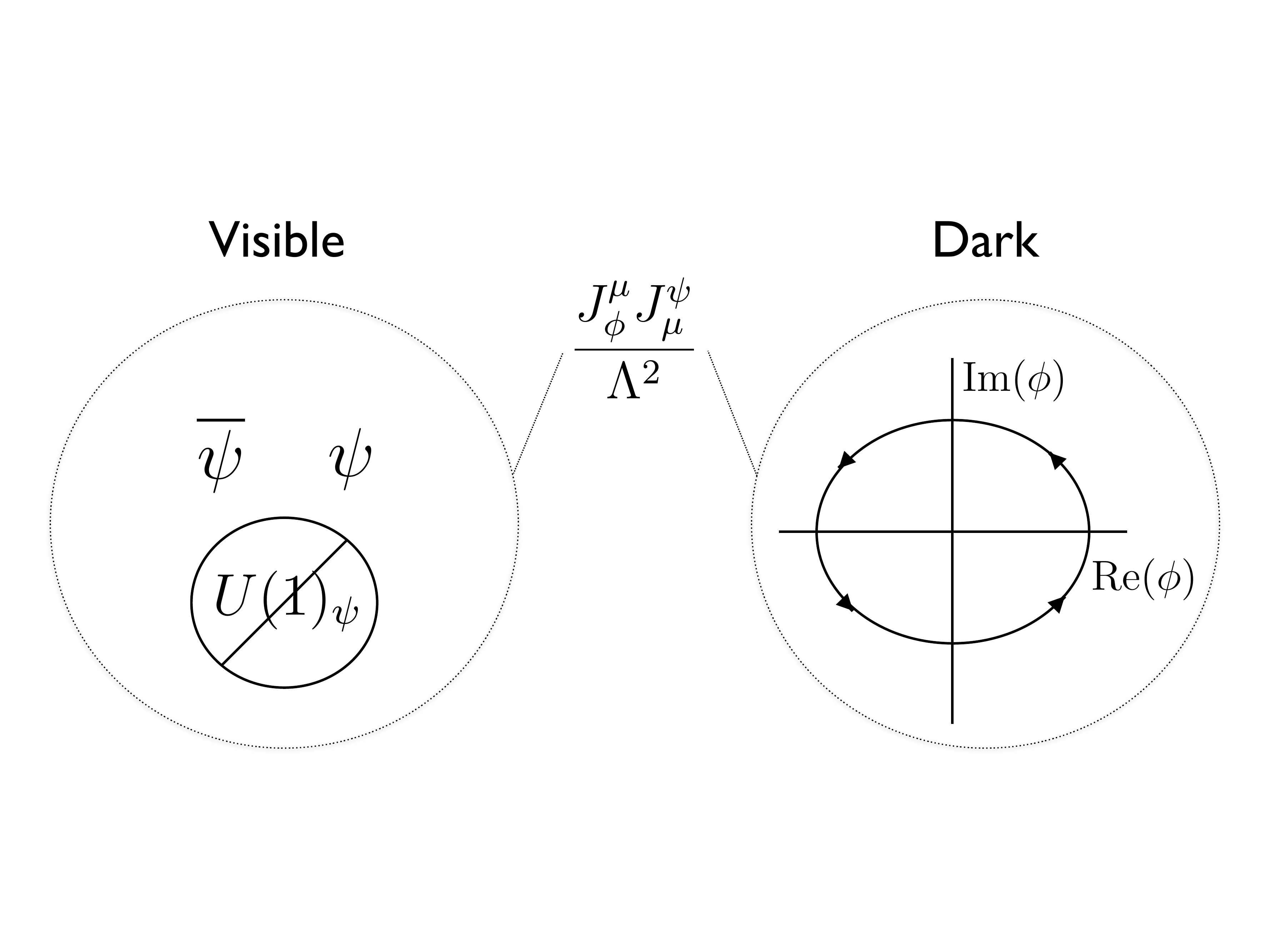}
\caption{A diagrammatic picture of our toy model.  The visible sector consists of a single fermion $\psi$ and an interaction that breaks $\psi$ number.  The dark sector consists of a single classical field $\phi$ which carries an asymmetry.  These two sectors communicate through a single higher dimensional current-current interaction.}
\label{Fig: Toy}
\end{center}
\end{figure}

In this section we present a simple toy example which exhibits all of the qualitative features of our mechanism.  We assume that the theory has two sectors that are not necessarily in thermal contact. We denote them as the dark sector and the visible sector.  The dark sector consists of a single complex field $\phi$.  The visible sector contains a single massless fermion $\psi$ and a $\psi$ number violating interaction that freezes-out at a temperature $T_\psi$.  The only interaction connecting the two sectors is the higher dimensional operator
\bea \label{Eq: higher dimensional}
\mathcal{L} \supset \frac{J^\mu_{\phi} J_{\mu}^{\psi}}{\Lambda^2} = i \frac{\left(\left(\partial^\mu \phi\right)^\dagger \phi- \phi^\dagger \partial^\mu \phi\right) \left ( \overline \psi \gamma_\mu \psi \right )}{\Lambda^2}\, .
\eea
The model is shown diagrammatically in Fig.~\ref{Fig: Toy}.  
For our mechanism to work, we need $J_\phi^0$ to acquire a background value.  To this end, we assume that $\phi$ is a classical field with an asymmetry and a potential dominated by its mass term. With this setting in mind we can study the evolution of the field after its asymmetry is generated,
\bea
\ddot \phi + 3 H \dot \phi + m_\phi^2 \phi =0\, .
\eea
We further make the simplifying assumption that $\phi$ is completely asymmetric.  Then in the limit $m_\phi \gg H$, the late time solution of the previous equation is
\be
\phi(t) \approx \phi_0 \left( \frac{a(t_0)}{a(t)} \right)^{-3/2} e^{i m_{\phi} t}\, ,
\ee
and we have a redshifting dark matter asymmetry $\langle J_\phi^0 \rangle = 2 m_\phi |\phi_0 |^2 \left ( a(t_0)/a(t) \right )^{3}$. To fix $\phi_0$, we assume that, as in the case of dark matter, we know the energy density $\rho_0$ of $\phi$ at some later time $t_0$, when the universe is at the temperature $T_0$.  In particular, we imagine that we have measured
\bea
\rho_0 = 2 m_{\phi}^2 |\phi_0 |^2\, .
\eea
Plugging this value for the amplitude of the background field into Eq.~(\ref{Eq: higher dimensional}), we find that the visible sector is modified in the following way,
\bea
\mathcal{L} \supset \frac{\rho_0}{m_\phi \Lambda^2} \frac{T^3}{T_0^3} \left ( \overline \psi \gamma_0 \psi \right ) \equiv  \mu \left ( \overline \psi \gamma_0 \psi \right ) = \mu\left(n_\psi -n_{\overline \psi}\right)
\eea
where we have traded the scale factor for the temperature.
We see that the effect of the asymmetry in the dark sector is to induce a splitting between $\psi$ and $\bar \psi$ energy levels; the background value of $\phi$ acts like a chemical potential for the $\psi$ field. In the presence of $\psi$ number violating interactions, thermal equilibrium carries an asymmetry in $\psi$
\bea
n_{\psi}-n_{\overline \psi} &=& 2\int \frac{d^3p}{(2\pi)^3} \left( \frac{1}{e^{(p-\mu)/T} + 1} - \frac{1}{e^{(p+\mu)/T} + 1} \right) \nn \\
&\approx& \left( \frac{\mu^3}{3 \pi^2}  + \frac{\mu T^2}{3} \right) \approx  \frac{\mu T^2}{3}, \quad T\gg \mu \, , \label{Eq: asymmetry}
\eea
where the additional factor of 2 comes from the two spins of $\psi$.  Notice that this asymmetry decreases with temperature.  At some temperature $T_\psi$, $\psi$ number violation freezes-out and the asymmetry can no longer change.  We find that the final $\psi$ number asymmetry is proportional to the $\phi$ field asymmetry
\bea
Y_\psi \equiv \frac{n_{\psi}-n_{\overline \psi} }{s} \approx \frac{n_{\overline \phi}-n_\phi}{s}\frac{T_\psi^2}{3 \Lambda^2} = Y_\phi\frac{T_\psi^2}{3 \Lambda^2}
\eea
where $s$ is the total entropy per comoving volume. If we know both the ratios of energy densities of $\psi$ and $\phi$ today and $T_\psi$, this relation gives the scale of the interaction $\Lambda$ between the two sectors as a function of $m_\phi$ and $m_\psi$. In the next section we show what values of $\Lambda$ lead to successful baryogenesis in the presence of different baryon number violating interactions.

The most interesting feature of this result is that the asymmetry in the visible sector can be generated by an asymmetry of a classical field $\phi$, without the need of thermal contact between the two sectors of the theory. Furthermore we show in Section~\ref{sec:constraint} that thermal contact is not needed even if $\phi$ is more accurately described by a thermal ensemble of particles.  This approach allows for two notable differences between our model and a traditional asymmetric dark matter scenario. First, we can make the dark sector extremely cold, as we did in this toy model by choosing a zero-momentum classical field.  Extremely cold dark sectors evade cosmological constraints on the dark matter mass and we can easily accommodate sub eV asymmetric dark matter. Second, we can easily have dark sectors with large asymmetric components of dark matter and no need to annihilate away the symmetric component.

\section{Baryogenesis from Selfish Dark Matter}\label{sec:chem}

In this section, we expand the toy model discussed in the previous one, constructing a theory that can generate the observed baryon asymmetry starting from an asymmetric dark sector.  As mentioned before, the dark matter asymmetry generates a chemical potential for the baryons through the higher dimensional current-current operator shown in Eq.~(\ref{eq:JJ}). Rather than discussing the operator directly, we introduce explicitly a $U(1)$ gauge boson in order to capture the effect of light mediators and high temperatures. To be concrete, we consider a specific model where $B \pm L+D$ is gauged and then spontaneously broken. Here $B(L)$ is baryon (lepton) number and $D$ the dark matter number.  A gauged anomalous symmetry implies the presence of a UV cutoff.  At the cutoff, there can either be new particles charged under $B+L$ and $SU(2)$ or the $B\pm L + D$ gauge boson could be a spin one meson of a confining sector.

We first discuss the case in which $n_D - n_{\overline D} = n_{DM}$ is a particle number asymmetry, then consider an asymmetric classical field.  In the following we always assume that dark matter number violation freezes-out at a temperature $T_D$ much larger than the temperature at which baryon or lepton number violating interactions fall out of equilibrium. 

\subsection{Particle Dark Matter}

We first consider the case where the dark matter is described by a set of particles. As mentioned before, we introduce a $B \pm L+D$ gauge interaction mediated by the vector boson $A_\mu$. The finite temperature Lagrangian for the $A_0$ component of the $B \pm L+D$ gauge field is
\be
\mathcal{L} = \frac{m_A^2(T)}{2} A_0 A^0 + g_A A_0\left( J^0_{D} + J^0_{B \pm L} \right) + \cdots \label{eq:UVsimple}
\ee
where
\bea
m_A^2(T) = m_A^2+\frac{g_A^2 T^2}{6} \sum_i q_i^2 g_i r_i^2\, . \label{eq:Dmass}
\eea
Here $r_i = T_i/T$ is the ratio of the temperature of particle $i$ to the temperature of the SM photons and the sum runs over all Weyl fermions and scalars charged under $B \pm L+D$ with mass less than $T_i$. The thermal contribution is the well known Debye mass for the electric field. In our convention $g=1$ for a Weyl fermion and $g=2$ for a Dirac fermion. 

In the presence of a dark matter asymmetry, $\langle J^0_D \rangle = n_{DM}$, the interaction in Eq.~(\ref{eq:UVsimple}) generates a tadpole for $A_0$ which gives it  a Lorentz-violating vacuum expectation value.  The non-zero vev of $A_0$ acts as a chemical potential for the Standard Model particles and  as long as $B \pm L$ violation is in equilibrium we have
\bea
\label{eq:nBL}
Y_{B \pm L} = \frac{a_\pm}{6 s} g_A \langle A_0 \rangle T^2 = -Y_{DM} \frac{a_\pm}{\sum_i q_i^2 g_i r_i^2 + a_\pm + \frac{6 m_A^2}{g_A^2 T^2}}\, .
\eea
If there are no new particles charged under $B$ or $L$ other than the SM fermions, the $a_{\pm}$ coefficients can be computed from the relations between SM chemical potentials (see Appendix~\ref{sec:asym}) giving $a_+ =408/37$ and $a_- = 79/11$. In spite of the intermediate step involving the $A_0$ vev, this is completely equivalent to the toy model described in the previous section. When the dark matter current gets a background value, the SM obtains a chemical potential $\mu$ through the gauge interaction that connects the two sectors. Here $\mu \sim \langle A_0 \rangle \sim n_{DM}$. Since we have decided to work in unitary gauge, the description in terms of the $A_0$ vev is more convenient. However we could have computed the matrix element for the current-current interactions connecting SM and dark sector and obtained the same result in a manifestly gauge invariant way.

We can see from Eq.~(\ref{eq:nBL}) that in the limit of large $m_A$, we recover the higher dimensional operator result with $\Lambda=m_A/g_A$.  In this limit, the baryon number density is smaller than the dark matter asymmetry, $Y_B\sim Y_{DM} g_A^2 T^2/m_A^2$.  In the opposite limit, $m_A/g_A \ll T$, the asymmetry in the dark matter and the visible sector are approximately equal, $Y_B\sim Y_{DM}$. In more detail, for a fixed dark matter asymmetry there is an upper bound on the asymmetry in the SM given by
\bea
|Y_{B \pm L}| \le \left|\frac{a_\pm}{\sum_i q_i^2 g_i r_i^2 + a_\pm} Y_{DM}\right|\, .
\eea 
In the minimal model where only the dark matter is charged under $U(1)_D$, only the Standard Model is charged under $U(1)_{B,L}$ and the temperature of the two sectors are the same, we have $|Y_{B \pm L}| \le (0.4 , 0.3) \, |Y_{DM}|$ for $B+L$ and $B-L$ violation, respectively. The inequality is saturated for $T_{B \pm L} \gg m_A/g_A$. This implies an upper bound on the dark matter mass needed to reproduce the observed baryon asymmetry of $\OO\left({\rm GeV}\right)$.  A priori there is no lower bound as long as $T^2_B/\Lambda^2$ can be made arbitrarily small. In the next section we discuss in more detail how to evade cosmological constraints on the dark matter mass and allow $m_{DM} \ll$~keV.

So far we have computed the $B\pm L$ asymmetries.  To relate them to the observed baryon asymmetry we need to take into account the effect of SM electroweak sphalerons and do a chemical potential calculation.  This is done in detail in Appendix~\ref{sec:asym} and in this section we just present the result. In the presence of $B+L$ violation only and having gauged $B+L+D$, from Eq.~(\ref{eq:nBL}) we have 
\bea
Y_{B} \approx -\frac{5 Y_{DM}}{\frac{6 m_A^2}{g_A^2 T^2_{B+L}}+26 }\label{eq:BpLY}
\eea
for temperatures below the electroweak phase transition. To obtain Eq.~(\ref{eq:BpLY}), we have assumed that the two sectors have the same temperature and that the contribution from the $B+L+D$ Higgs to the Debye mass of $A_0$ can be neglected. If the last source of $B+L$ violation to freeze out is the electroweak sphalerons, i.e. $T_{B+L}\approx 132$~GeV~\cite{D'Onofrio:2014kta}, we can solve Eq.~(\ref{eq:BpLY}) for $m_A/g_A$ and obtain the scale of the interaction needed for successful baryogenesis. To do this we need to specify what fraction of the current dark matter energy density is due to the asymmetry. We introduce
\bea \label{Eq: r}
R_D = \left|\frac{n_D - n_{\overline D}}{n_D + n_{\overline D}}\right|
\eea
so that $R_D=0$ corresponds to symmetric dark matter while $R_D=1$ corresponds to completely asymmetric dark matter. Then today $Y_{B}/Y_{DM}=(m_{DM}/m_p)(\rho_{B}/R_D \rho_{DM})\approx (m_{DM}/5 R_D m_p)$. This allows us to solve for the strength of the interaction and we obtain
\bea
\frac{g_A}{m_A} \approx 4 \times 10^{-3}\; {\rm GeV}^{-1} \frac{1}{\sqrt{\left(\frac{R_D \;{\rm GeV}}{m_{DM}}\right)-1}}\, .\label{eq:B+Lfinal}
\eea
In this case, the upper bound on the dark matter mass discussed before is approximately 1 GeV times the fraction of the asymmetric component of dark matter. Not surprisingly the strength of the interaction is set by the $B+L$ violation freeze-out temperature $m_A/g_A\sim T_{B+L}$ and the dark matter mass.

If $B-L$ violation is present and decouples at a temperature $T_{B-L} \gg T_S$, combining Eq.~(\ref{eq:nBL}) and the results in Appendix~\ref{sec:asym} we get
\bea
Y_{B}(T_{B-L}) =  -\frac{948}{407}\frac{Y_{DM}(T_{B-L})}{\frac{6 m_A^2}{g_A^2 T^2_{B-L}}+\frac{244}{11}}+\OO(T_S^2/T_{B-L}^2)\label{eq:BmLY}\, .
\eea
So to reproduce the observed baryon asymmetry we find that
\bea
\frac{g_A}{m_A} \approx \frac{0.5}{T_{B-L}} \frac{1}{\sqrt{\left(\frac{0.5 R_D\;{\rm GeV}}{m_{DM}}\right)-1}}\, \label{eq:B-Lfinal}
\eea
is needed, giving the upper bound $m_{DM}\lesssim 0.5 R_D$~GeV.

While not discussed in detail in this paper, this approach can also be used assuming that a SM asymmetry is generated first and then used to generate the dark matter asymmetry.  In this case, we have 
\bea
Y_{DM} = -Y_{B \pm L} \frac{b}{\sum_i q_i^2 g_i r_i^2 + b + \frac{6 m_A^2}{g_A^2 T^2}}
\eea
where $b$ is the equivalent of $a_\pm$ for the dark matter sector.  Using the representative value of $b = 2$, assuming that the two sectors are at the same temperature and taking the temperature at which DM number violating interactions freeze-out to be $T_D$, we find that
\bea
\frac{g_A}{m_A} \approx \frac{0.6}{T_{D}} \frac{1}{\sqrt{\frac{m_{DM}}{44\;{\rm GeV}}-1}}
\eea
In this scenario, the dark matter mass has a lower bound of $\approx 44$ GeV.  Thus we see that selfish asymmetric dark matter can be either extremely light or extremely heavy compared to the proton. This behavior is easy to understand in the two extreme cases.  For $m_A \gg T_{D}$ only a small asymmetry is generated in the dark matter sector and to obtain $\rho_{DM}/\rho_B \approx 5$, we need $m_{DM}\gg m_p$. The lower bound is saturated in the opposite limit where $n_{DM}\sim n_{B\pm L}$ and $\OO(1)$ factors explain the difference between DM and proton masses. 

\subsection{Classical Dark Matter}\label{sec:codm}

As mentioned before, the mechanism has the potential to accommodate extremely light asymmetric dark matter. The simplest way to exploit this feature is to have the dark matter energy density in the form of zero momentum oscillations of a classical field, as shown in our toy example. In this setting, light dark matter automatically evades Big Bang Nucleosynthesis and structure formation bounds. A well known example is that of a misaligned axion field. In our setup we need a particular form of the classical field reminiscent of squarks or sleptons in Affleck-Dine baryogenesis~\cite{Affleck:1984fy}, since we would like the operator
\be
J_{DM}^0=i\left(\left(\partial^0 \phi\right)^\dagger \phi-\phi^\dagger \partial^0 \phi\right)
\ee
to acquire a background value. This is easily achieved by assuming that the oscillations of the DM field are in the form
\be \label{Eq: classical}
\phi(t)=\phi_0 e^{i m_{DM} t}\, .
\ee
For simplicity we assumed, as in the toy model in Section~\ref{Sec: Toy}, that terms other than the quadratic mass term in the $\phi$ potential are negligible and hidden the effects of the expansion of the universe inside $\phi_0$.  We have also taken the classical dark matter field to be completely asymmetric.  Following the steps in Section~\ref{Sec: Toy}, we have that at late times $\rho_{DM}= 2 m_{DM}^2 \phi_0^2$ and
\be
\phi_0= \sqrt{ \frac{\rho_{DM}}{2 m_{DM}^2} }\, ,
\ee
where $\rho_{DM}$ redshifts as the energy density of a non-relativistic particle. With these choices, the dark matter current acquires the background value $\langle J_{DM}^0 \rangle = n_{\overline \phi}-n_{\phi}=2 m_{DM}\phi_0^2$. From this point onwards the discussion is completely equivalent to the particle dark matter case. The only difference is that we have to include a contribution to the gauge boson mass proportional to the dark matter energy density
\be
\LL \supset g_A^2 Q_{DM}^2 A_\mu A^\mu |\phi(t)|^2 = \frac{1}{2} g_A^2 Q_{DM}^2 \frac{\rho_{DM}}{m_{DM}^2} A_\mu A^\mu\, ,
\ee
where we have allowed for the possibility of gauging a quantity proportional to dark matter number and giving the dark matter a charge $Q_{DM}$.
Repeating the steps in the previous subsection in the presence of this extra mass term, we obtain for $B+L$ violation from electroweak sphalerons
\bea
\frac{g_A}{m_A} \approx 4 \times 10^{-3}\; {\rm GeV}^{-1} \frac{1}{\sqrt{Q_{DM}\left(\frac{\;{\rm GeV}}{m_{DM}}\right)-Q_{DM}^2\left(\frac{6\times10^{-4}\;{\rm GeV}}{m_{DM}}\right)^2-1}}\, ,\label{eq:B+Lfinalscalar}\, 
\eea
The result has not changed considerably with respect to particle dark matter, but in this case we have a lower bound on the dark matter mass: $0.6\;{\rm keV} \lesssim m_{DM}/Q_{DM} \lesssim$~GeV. In contrast to the previous case, in order for the dark matter to be extremely light, its charge must be very small. This arises naturally in models where the dark matter obtains a charge under the $B\pm L$ gauge group through kinetic mixing. Thus a lighter dark matter particle is also less likely to be in thermal contact with the SM.

In the presence of $B-L$ violation at high temperatures, we have instead
\bea
\frac{g_A}{m_A} \approx \frac{0.5}{T_{B-L}} \frac{1}{\sqrt{Q_{DM}\left(\frac{0.6 \, \;{\rm GeV}}{m_{DM}}\right)-Q_{DM}^2\left(\frac{7\times 10^{-9}\;{\rm GeV}}{m_{DM}}\frac{T_{B-L}}{m_{DM}}\right)-1}}\, .\label{eq:B-Lfinalscalar}
\eea
In this setup it might seem that there is a lower bound on the DM mass from the requirement that $\phi$ starts oscillating before baryon number violation freezes-out. However from the equations of motion for $\phi$ and $J_{DM}^\mu$ we find that the only lower bound comes from structure formation. The classical field evolves as 
\be
0=\ddot{\phi}+3 H\dot{\phi}+\frac{\partial V}{\partial \phi}\approx \ddot{\phi}+3 H\dot{\phi}+m_{DM}^2 \phi\, ,
\ee
where for simplicity we have assumed that $\phi$ does not decay into pairs of gauge bosons or SM particles. In the limit $H \gg m_{DM}$, this equation has the solution
\bea
\phi(t) = \phi_1 - \frac{2 \phi_2 }{\sqrt{t}}\, ,
\eea
where we have assumed radiation domination.  The conserved asymmetric number density is
\bea
J^0_{DM}(t)= i \frac{(\phi_1^\dagger \phi_2 - \phi_2^\dagger \phi_1)}{t^{3/2}}\, ,
\eea
which is non-zero even before $\phi$ starts oscillating sinusoidally. This shows that once the asymmetry is generated it just redshifts according to its equation of motion
\be
\frac{d J^0_{DM}}{dt}=-3 H J^0_{DM}\, ,
\ee
regardless of whether $H> m_{DM}$ or $H< m_{DM}$. Unlike the case of the axion, the time dependent piece is critical when $H > m_{DM}$. In the opposite limit, $m_{DM} \gg H$, the solution is given by Eq.~(\ref{Eq: classical}) with $\phi_0 \propto t^{-3/4}$ and the equation of state is $w \approx 0$. As we have just seen, we do not need this regime to set in when the asymmetry is communicated to the visible sector. However in order not to disrupt matter radiation equality at $T_{EQ}\approx$~eV, we need the classical field to behave as cold dark matter, i.e. $m_{DM}\gtrsim H(T_{EQ}) \approx 10^{-28}$~eV. 

Note that here we have not specifically used the Affleck-Dine (AD) mechanism to generate the asymmetry in the dark sector. In AD models the field $\phi$ acquires a minimum away from zero due to supersymmetry breaking effects proportional to Hubble and then spirals back towards the origin as $H$ decreases. The final value of the asymmetry is reached when the soft supersymmetry breaking mass $m_\phi$ becomes larger than Hubble and the field starts oscillating. So we can use the results in this section starting from an AD setup only if $m_{DM} > H(T_{B\pm L})$. For the elctroweak sphaleron case, this implies $m _{DM} \gtrsim 10^{-5}$~eV. We leave to future work the exploration of the interesting possibility of an Affleck-Dine model where $m_{DM} < H(T_{B\pm L})$. 

\subsection{The Dark Matter Mass}
In this section we have shown that, starting from an asymmetry in the dark sector, we need an interaction between the SM and the dark matter whose scale is roughly set by the temperature at which baryon number violation freezes-out $\Lambda = m_A/g_A \sim T_B$. The strength of the interaction decreases with the dark matter mass as $g_A/m_A \sim (m_{DM}/{\rm GeV})^{-1/2}$. This is not surprising since in our setup the number asymmetry in the visible sector is proportional to the dark matter number asymmetry.  At a fixed DM energy density, a lighter dark matter particle corresponds to a larger number asymmetry and thus a less efficient generation of the asymmetry. This also implies an upper bound on the dark matter mass of $\OO({\rm GeV})$ corresponding to the case in which the interaction is strong enough to give $n_B \approx n_{DM}$. 

We have discussed the generation of the asymmetry both for particle dark matter and an asymmetric classical field. The two cases are very similar with the exception of a lower bound on the dark matter mass in the classical case, which can be made sub eV, by reducing the dark matter charge. A second bound on the dark matter mass, common to the particle and classical field cases, arises from the requirement that the vacuum energy stored in the background value of $J^0_{DM}$, or equivalently the $A_0$ vev, does not dominate the equation of state of the universe when $B\pm L$ violation freezes-out\footnote{We thank Kfir Blum for pointing this out to us.}. Since our vacuum energy redshifts faster than radiation and matter, this insures that it will not dominate at later times. We find that if the dark matter mass is too small, its number density in the early universe is large, leading to a sizable vacuum energy $\rho_\Lambda \sim \langle A_0 \rangle^2\sim n_{DM}^2$. Therefore we can obtain a lower bound on the dark matter mass: $m_{DM}\gtrsim c_{B\pm L} \times 10^{-9}\;{\rm eV} \rho_R(T_{B\pm L})/\rho_\Lambda(T_{B\pm L})$, where $\rho_R$ is the radiation energy density and $c_{B+L}\approx 2$, $c_{B-L}\approx 4$. If we ask that $\rho_\Lambda(T_{B\pm L})\approx 0.1 \rho_R(T_{B\pm L})$, for completely asymmetric dark matter this gives the strongest constraint on the dark matter mass so far, implying $m_{DM}\gtrsim 10^{-8}$~eV. For fermionic dark matter this is superseded by a bound arising from dark matter density measurements today. If fermionic dark matter was too light, the Pauli exclusion principle would not allow for the densities that have been observed in dwarf galaxies.  Thus fermionic dark matter must have a mass larger than $\approx 0.5$ keV~\cite{Tremaine:1979we,Lin:1983vq}. The bound can vary by factors of a few depending on the assumptions on the dark matter phase space distribution~\cite{Boyarsky:2008ju, Domcke:2014kla} and can be affected by dark matter self-interactions. Rather than discussing these details, we take the order of magnitude estimate.  Summarizing our findings, for the mechanism to produce the observed baryon asymmetry we need 
\bea
&{\rm \bf Fermions}& \quad \OO(0.5\;{\rm keV}) \lesssim m_{DM} \lesssim R_D \; {\rm GeV}\, , \nn \\
&{\rm \bf Bosons}& \quad 10^{-8} \;{\rm eV}\; \frac{R_D}{\epsilon+R_D^2} \lesssim m_{DM} \lesssim R_D \; {\rm GeV}\, , \nn
\eea
where $R_D\in [0,1]$, $\epsilon\approx 10^{-18}$ and we do not show the small $\OO(1)$ differences in the bounds between $B+L$ violation only and $B-L$ violation.

The results in this section allow for two important phenomenological consequences that we explore in the next two sections. First, we can obtain successful baryogenesis from extremely light asymmetric dark matter. This is a rather bleak scenario as far as detection goes since, as stated above, the typical interactions would be unobservably small: $g_A/m_A \sim (m_{DM}/{\rm GeV})^{-1/2}$. However it is  a conceptual possibility of which our models are the only known example in the literature. We discuss in more detail how we can accommodate light dark matter in Section~\ref{sec:constraint}. The second interesting feature of the mechanism is an Asymmetric WIMP miracle.  In its minimal realization, electroweak sphalerons are the only source of baryon number violation while both the dark matter and the new gauge boson have GeV scale masses.  This simple realization can give signals at direct detection experiments and colliders.
We explore this possibility in Section~\ref{sec:sphalerons}.

\section{Absence of Thermal Contact} \label{sec:constraint}

An interesting feature of this model is that the dark matter does not need to be in thermal contact with the SM in order to generate a baryon asymmetry.  The reason for this surprising fact is that we use the presence of the asymmetric background number density of dark matter to generate the asymmetry in the SM.  As discussed in the previous section, the most familiar case of a background causing a non-trivial effect comes from coherent classical excitations of bosons.  For example, if the dark matter asymmetry was generated by Afleck-Dine baryogensis, then one expects it to be in the form
\bea
\phi_{DM}(t) = \phi_0 e^{i m_{DM} t}
\eea
which describes a coherent state of anti-particles with no particles present.  Because a coherent state is made from a large number density of particles, it is expected that it can have a non-trivial effect on the SM even if the interactions connecting the two sectors are too weak to maintain thermal equilibrium.

It is significantly less obvious how an asymmetry in the form of a number density of fermions or a non-coherent excitation of bosons can still affect the SM in the presence of very weak interactions. The resolution is that even if there is no thermal contact between the two sectors, the dark matter sector still affects the SM through loops. 

To illustrate this phenomenon in detail, we consider the case of fermionic dark matter.  For simplicity we use again the effective operator description and assume that the following is the only coupling between the visible and dark sectors
\bea
\mathcal{L} \supset \frac{\overline \psi_{DM} \gamma_\mu \psi_{DM} \overline \psi_{SM} \gamma^\mu \psi_{SM}}{\Lambda^2} \, .\label{Eq:higher}
\eea
We also assume that the temperatures of the SM and the DM sectors are small enough that this higher dimensional operator has frozen out.  As discussed before, if there is an asymmetry in the dark matter sector, then evaluating this operator on the background gives a chemical potential for the SM fields $\mu \sim n_{DM} / \Lambda^2$.  We now discuss an alternative way to view how the dark matter asymmetry generates this operator and show that it is explicitly independent of thermal contact between the two sectors.

Despite the fact that scattering between the two sectors has frozen out, Eq.~(\ref{Eq:higher}) still has loop level effects on the SM.  We consider the simplest such diagram where we close the loop of $\psi_{DM}$ particles generating an operator proportional to $\overline \psi_{SM} \gamma^\mu \psi_{SM}$:
\bea
\delta\mathcal{L} = \overline \psi_{SM} \gamma^0 \psi_{SM} \frac{T_{DM}}{\Lambda^2} \sum_n g_i \int \frac{d^3 \vec{p}}{(2 \pi)^3} \frac{\text{Tr} \gamma_0 \slashed{p}}{p^2-m_{DM}^2}\, .
\eea
The momentum in thermal field theory takes the form
\bea
p = (i \pi T_{DM} (2 n + 1) + \mu_{DM}\, , \,  \vec{p} )\, ,
\eea
Here $\mu_{DM}$ is the chemical potential that parametrizes the non-zero asymmetry in the dark matter sector, $n_{DM} \approx g_i \mu_{DM} T_{DM}^2/6$. We have accounted for the presence of this chemical potential in the dark matter equation of state in the standard way. We now make the simplifying assumptions that the dark matter is massless and that the chemical potential is small compared to $T_{DM}$.  In this limit, the correction to the Lagrangian is 
\bea
\delta\mathcal{L} = \overline \psi_{SM} \gamma^0 \psi_{SM} \left ( \frac{g_i \mu_{DM} T_{DM}^2}{6 \Lambda^2} + \frac{T_{DM}^3}{\Lambda^2} \mathcal{O}\left(\frac{\mu^3_{DM}}{T_{DM}^3}\right) \right ) = \frac{n_{DM}}{\Lambda^2}  \overline \psi_{SM} \gamma^0 \psi_{SM}
\eea
We have now arrived back at the result that the chemical potential for the SM is proportional to the asymmetric number density of dark matter.  However, this derivation should make clear the fact that the way this operator is generated has nothing to do with thermal contact between the two sectors.  Dark matter's asymmetric number abundance makes an appearance due to the fact that the dark matter propagator at finite density is modified from the vacuum propagator by the Pauli exclusion principle for fermions and Bose enhancement for bosons.  These factors are present regardless of whether or not the two sectors are in thermal equilibrium with each other.  Thus our effect is present even if the dark matter and the SM are not in thermal contact. 

\subsection{The Temperature of the Dark Sector}

The fact that this mechanism works without thermal contact is very important in opening up a large amount of parameter space.  If dark matter was in thermal equilibrium with the SM at any point, then Big Bang Nucleosynthesis (BBN) and structure formation would place strong constraints on its mass.  However, since our mechanism allows for the dark matter to be extremely cold, both constraints can be completely avoided.  To have a sufficiently cold dark sector relative to the visible sector, we can  have for example a zero-momentum classical dark matter field, as discussed in Section~\ref{sec:codm}. Even if BBN and structure formation constraints can be evaded, we briefly summarize them here. In the case of BBN, new relativistic degrees of freedom present at $T_{SM}\sim$  MeV affect the value of the Hubble parameter that in turn determines the primordial light elements abundances. This constraint can be applied by requiring that at 95$\%$ C.L. $\Delta N_{\rm eff} \lesssim 1$~\cite{Cooke:2013cba}, where
\be
\Delta N_{\rm eff} = \frac{\rho_X}{\rho_\nu} = a_X \left(\frac{T_{DM}}{T_\nu}\right)^{4} . \label{eq:BBN}
\ee
Here $\rho_{\nu(X)}$ is the neutrino (a generic dark sector particle $X$) energy density, $T_\nu$ ($T_{DM}$) is the neutrino (dark sector) temperature and $a_X=n_F(8n_B/7)$ for fermions with $2 n_F$ degrees of freedom (bosons with $2 n_B$).  The second equality is valid for $T_{DM} \gg m_X$. In the absence of thermal contact, $T_{DM}$ can be considered to be a free parameter. Therefore the bound can be completely evaded if the dark sector is not in thermal contact with the SM before BBN and was reheated to a smaller temperature, or if it was in thermal equilibrium with the SM but has decoupled before $T\approx 1$~MeV and cooled faster. It was shown that it is easy to accommodate MeV scale dark sectors as large a one generation version of the SM starting from the same reheating temperature~\cite{Feng:2008mu}. 

At a later epoch, warm dark matter particles can suppress density fluctuations on scales smaller than their free-streaming length. Measurements of the matter power spectrum include data from CMB experiments, galaxy surveys and the Lyman-$\alpha$ forest~\cite{Croft:2000hs, Gnedin:2001wg} and are in agreement with the $\Lambda$CDM model. To estimate the constraint on the dark sector temperature imposed by these experiments, we require conservatively that the dark matter free-streaming length $\lambda_{FS}=\int dt (v_{DM}(t)/a(t))$ be smaller than the scale probed by the Lyman-$\alpha$ forest measurements, i.e. $\lambda_{FS} \lesssim $ Mpc~\cite{Tegmark:2002cy}.  If the dark sector decouples when it is non-relativistic or is only reheated to non-relativistic temperatures, the free-streaming length grows logarithmically until matter radiation equality: $\lambda_{FS}\approx 10\;{\rm pc}\;\sqrt{1/g_{\star}(T_d)}\; (10\;{\rm MeV}/T_d)\sqrt{T_{DM}^d/m_{DM}}\log(t_{EQ}/t_d)$~\cite{Loeb:2005pm} where $T_d$ ($T_{DM}^d$) is the temperature of the visible (dark) sector at kinetic decoupling.  If the two sectors were in thermal contact, these temperatures are the same.  In absence of thermal contact at any point in the history of the universe, $T_{DM}^d$ is a free parameter corresponding to the dark sector temperature when the SM is at the temperature $T_d$. Requiring $\lambda_{FS} \lesssim $ Mpc gives a mild constraint on the ratio $T_d^{DM}/m_{DM}$. Even when the free-streaming length is small the dark sector can not decouple too late, because the imprint of acoustic oscillations of the radiation fluid on the DM would suppress structure on scales $\propto 10\; {\rm pc}\; (10\;{\rm MeV}/T_d)$~\cite{Loeb:2005pm}. If the dark matter decouples when relativistic, the free-streaming length is
\be
\lambda_{FS} \approx 0.3\;{\rm Mpc} \frac{{\rm keV}}{m_{DM}}\frac{T_{DM}}{T_{NR}}\frac{1}{\sqrt{g_\star(T_{NR})}}\left[\log(t_{EQ}/t_{NR})+2\right]\, ,
\ee
where we have assumed that the DM becomes non-relativistic at $t_{NR} < t_{EQ}$, when the SM temperature is $T_{NR}$ and the dark sector temperature $T_{DM}$. We have also let the DM free-stream with $v_{DM}=c$ from $t=0$ to $t_{NR}$. If we require $\lambda_{FS} \lesssim $ Mpc for a particle becoming non-relativistic right at matter radiation equality we obtain $T_{DM} \lesssim 10^{-2}\; m_{DM}$. Note that if all particles in the dark sector are approximately at the scale $m_{DM}$, once we take $T_{DM} \lesssim 10^{-2}\; m_{DM}$, the bound on $N_{\rm eff}$ from CMB measurements~\cite{Planck:2015xua} is easily satisfied. It is also worth mentioning that constraints arising from energy injection in the CMB~\cite{Planck:2015xua} or dark matter self-interactions~\cite{Buckley:2009in} that are typically important for light dark matter do not reduce the parameter space of the models that we consider here. This is due to the fact that the strength of the dark matter interactions is not related to its mass and in this work we have theories where $m_A/g_A \gtrsim  100\; {\rm GeV}\gg m_{DM}$.

Finally, we note that the gauge boson connecting the two sectors can potentially bring them in thermal equilibrium.  For example, if we are using the $B+L$ violation provided by sphalerons to generate the asymmetry (i.e. Eq.~(\ref{eq:B+Lfinal}) or~(\ref{eq:B+Lfinalscalar}) set the strength of the new interaction) and the higher dimensional operator approximation, then if $m_{DM} \gtrsim 0.06$ keV, the interaction will not have frozen out by the time sphalerons become inactive. However if we make the dark photon sufficiently light at fixed $g_A/m_A$ these interactions become IR dominated and never thermally couple the two sectors. As we can easily work in the regime where the two sectors never approach thermal equilibrium, we do not give the results of a complete computation as it would involve a detailed freeze-out or freeze-in calculation~\cite{Hall:2009bx}.

\section{Phenomenology}
\label{sec:models}\label{sec:UVBmL}

In the previous sections we have described the generation of the baryon asymmetry, deriving a range of viable dark matter masses. Here we discuss two realizations of the general mechanism with interesting phenomenological consequences. We first introduce a model where the SM asymmetry is dominated by lepton number violation in the early universe from Majorana neutrino masses. In the second model, we generate the asymmetry through SM electroweak sphalerons and find observable signals at present and future colliders and direct detection experiments.

The first model is extremely simple and can be summarized in a few lines. Once we gauge the $B-L+D$ symmetry, a single source of spontaneous breaking of lepton number can be used to generate left-handed neutrino masses and the baryon asymmetry at the same time. In more detail, we assume that in the UV both $B-L$ and $D$ are good symmetries of the theory.  In order to remove the gravitational anomaly of $B-L$, we introduce right handed neutrinos $\nu_R$.  A new Higgs field, $\phi_{B-L+D}$, neutral under all SM gauge groups and with $L=2$, provides the spontaneous breaking of lepton number giving a mass to right-handed and left-handed neutrinos through the Yukawa interactions
\bea
\delta \LL = -y_1 \phi_{B-L+D} \nu_R \nu_R - y_2 H l \nu_R + c.c. 
\eea
where $H$ is the SM Higgs and $l$ the SM left-handed lepton doublet. If we take $\langle \phi_{B-L+D}\rangle =v_{B-L+D}\approx 10^{14}$~GeV and assume that $y_{1,2}$ are $\OO(1)$, we obtain light neutrino masses from the see-saw mechanism in the right ballpark $m_{\nu} \sim y_2^2 v^2/y_1 v_{B-L+D} \approx 0.1$~eV. At the same time, we have generated lepton number violating interactions\footnote{After integrating out the right-handed neutrinos, we obtain the dimension five operator discussed in Section~\ref{sec:BmL}.} that freeze-out at a temperature
\bea
T_L \sim \frac{y_1^2 m_A^2}{y_2^4 g^2_A M_{Pl}}\approx 10^{10}\;{\rm GeV}\, .
\eea
Using Eq.~(\ref{eq:B-Lfinal}), we see that this model requires a dark matter mass of $\approx$20 eV.  This model has the appealing feature of generating neutrino masses and the baryon abundance from the same source.  However it gives no observable signals at present or planned experiments (see for instance~\cite{Essig:2013lka}). In the next subsection, we focus on a model with a much richer phenomenology.
 
\subsection{Baryogenesis from Electroweak Sphalerons}\label{sec:sphalerons}
If the generation of the asymmetry is due to electroweak sphalerons and the dark matter mass is not too far removed from the GeV scale, it is natural to expect the $B+L+D$ photon to be observable and the dark matter to have a direct detection cross section in the ballpark of a WIMP, since as we showed before, $m_A/g_A \approx T_S \approx 100$ GeV. In this section we discuss possible signals and comment on the constraints on our parameter space. For simplicity we assume that the dark and visible sectors have been in thermal contact at least until $T_S$, that dark matter today is mostly asymmetric $R_D\approx 1$ and consider only particle dark matter.  

Before discussing the phenomenology of the model, it is worth pointing out an indirect bound on the dark matter mass specific of this realization of the mechanism. The theory requires gauging the $B+L+D$ symmetry, so new matter must be added to cancel the $B+L$ anomalies\footnote{The $B+L$ symmetry has two anomalies, $\Tr \big ( SU(2)^2 \times U(1)_{B+L} \big ) = \mathcal{A}_{221} = 3$ and $\Tr \big ( U(1)_Y^2 \times U(1)_{B+L} \big ) = \mathcal{A}_{YY1} = 3$.}. We do not add matter charged under $D$ since it would make the $U(1)_D$ symmetry anomalous and change the dark matter asymmetry. The expectation is that the new particles appear to restore unitarity around the scale $m_A/g_A$
\bea
\Lambda_{NP} \lesssim 4\pi\frac{m_A}{g_A} \approx 3.3\;\text{TeV}\sqrt{\left(\frac{1.16\;{\rm GeV}}{m_{DM}}\right)-1}\, .
\eea
Depending on whether or not these new particles are charged under $SU(3)_c$, current collider bounds can go anywhere between $100$~GeV and $1$~TeV. However since we still want sphalerons to violate $B+L$ around $T_S$ we require conservatively that $\Lambda_{NP}> 1$~TeV and we obtain $m_{DM} < 1.06$~GeV\footnote{Note that for an abelian symmetry, using charges that are orders of magnitude away from each other the appearance of new physics can be delayed up to $\Lambda_{NP} \lesssim \frac{64 \pi^3 m_A}{g_A g_{2,Y}^2 \mathcal{A}_{221,YY1}} \approx 10^4 \, \text{TeV} \sqrt{\frac{\text{GeV}}{m_{DM}}-1}$~\cite{Preskill:1990fr}, but we do not consider this possibility here.}. This might seem a trivial extension of $m_{DM} < 1.16$~GeV, but it strongly reduces the allowed values of the gauge coupling for a given $B+L+D$ photon mass. Before $g_A$ was bounded from above only by a perturbativity requirement, for every $m_A$, while now $g_A < 4\pi m_A/\Lambda_{NP} \lesssim 4\pi \times 10^{-3}\; m_A/{\rm GeV}$. This still leaves open a large parameter space. As we have seen in Section~\ref{sec:chem}, baryogenesis fixes only $g_A/m_A$:
\be
g_A \approx 4 \times 10^{-3} \left(\frac{m_A}{\rm GeV}\right)\frac{1}{\sqrt{\left(\frac{{\rm GeV}}{m_{DM}}\right)-1}}\, ,\label{eq:gABpL}
\ee
and the mass of the new gauge boson can span over many orders of magnitude, giving very different signatures. Additionally, the mechanism works even if only $B+D$ or $L+D$ are gauged and the bounds change drastically between different choices of gauge group. Therefore in what follows we give only a rough overview of the phenomenology, focusing on possible signals in laboratory experiments. 

If baryon number is gauged, we have an observable direct detection cross section for dark matter masses not too far from the GeV scale. The leading term in the single nucleon cross section for both fermions and bosons reads
\be
\sigma_N=\frac{\mu^2_N g_A^4}{\pi m_A^4}\approx \frac{10^{-38}\;{\rm cm^2}}{\left(1+\frac{\rm GeV}{m_{DM}}\right)^2\left(1-\frac{\;{\rm GeV}}{m_{DM}}\right)^2} \, , \label{eq:DD}
\ee
where $\mu=m_N m_{DM}/(m_N + m_{DM})$ is the usual reduced mass of the DM-nucleon system. As stated above, the requirement of generating the correct baryon asymmetry fixes $g_A/m_A$ as a function of $m_{DM}$. Thus, once we fix the dark matter charge (here $Q_{DM}=1$), the cross section depends only on its mass. This is strictly true only for $m_A \gtrsim q$, where $q=10^{-3} \mu_{N}$ is the relevant momentum transfer. For very light gauge bosons $m_A \ll {\rm MeV}/[1+({\rm GeV}/m_{DM})]$, the cross section is suppressed by powers of $m_A^2/q^2$ and acquires an explicit dependence on $m_A$. 

\begin{figure}[!t]
\begin{center}
\includegraphics[width=0.6\textwidth]{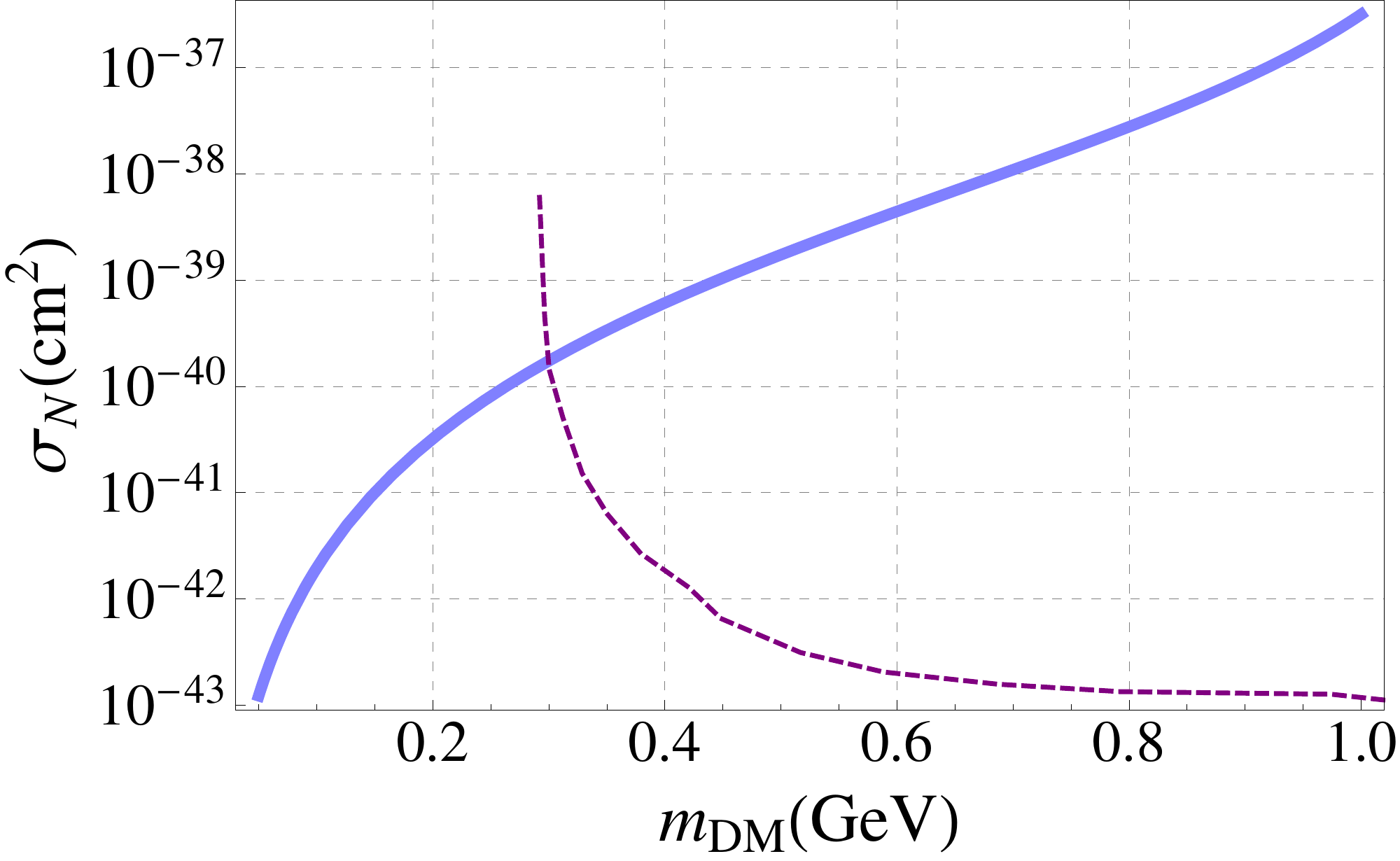}
\caption{Dark matter-nucleon scattering cross section as a function of dark matter mass (light blue). The purple dashed line shows the projected reach of SuperCDMS SNOLAB~\cite{Cushman:2013zza}}
\label{fig:DD}
\end{center}
\end{figure}

In Figure~\ref{fig:DD}, we plot the cross section shown in Eq.~(\ref{eq:DD}) and compare it with the reach of SuperCDMS SNOLAB~\cite{Cushman:2013zza}. The SuperCDMS upgrade has the potential to probe our parameter space up to $m_{DM}\approx 300$~MeV. For lower masses, the bound from Xenon10 on dark matter scattering off atomic electrons~\cite{Essig:2012yx} is not relevant in our parameter space. Also proposed direct detection electron scattering experiments ~\cite{Essig:2011nj} do not have the necessary sensitivity to probe this model. Even in the presence of a direct coupling to electrons our parameter space is out of reach for this proposal. Furthermore current direct detection experiments sensitive to low mass WIMPS such as DAMIC~\cite{Barreto:2011zu}, CDMSLite~\cite{Agnese:2013jaa}, LUX~\cite{Akerib:2013tjd}, Xenon10~\cite{Angle:2011th} and SuperCDMS~\cite{Agnese:2014aze} do not place constraints on dark matter masses below a GeV. Note that neutron star bounds on scalar asymmetric dark matter could already exclude the range of interest for direct detection~\cite{McDermott:2011jp}. Since the dark matter could be captured and reach densities large enough to create a black hole during the neutron star lifetime. However in our models, there is a strong self repulsion that prevents the dark matter from clumping into high enough densities and we do not have significant constraints in the mass range of interest for SuperCDMS SNOLAB~\cite{Bell:2013xk}.

After having discussed direct detection, we can now turn to the dark photon phenomenology. Depending on the new gauge boson mass and on whether or not lepton number is gauged, we can have signals also at the LHC and at $B$-factories. Even though this new gauge boson couples to $B$, $L$ or both, the constraints and the possible signals are very similar to those of a dark photon that obtains couplings to the SM through kinetic mixing.  As such, we simply refer to the new gauge boson as the dark photon. When discussing its phenomenology, we consider couplings given by dark matter masses between a keV and a GeV. Some regions of the parameter space with masses below an MeV are excluded by BBN. However changing the lower limit on the dark matter mass does not change the qualitative picture of the dark photon phenomenology that we are interested in, so we always consider the larger dark matter mass range.

For $m_A \gtrsim 10$~GeV, measurements of Drell-Yann (DY) production at the LHC are sensitive to couplings in the ballpark of $g_A \sim 10^{-2} \sqrt{10\%/{\rm BR}(A\to l^+l^-)}$ \cite{Hook:2010tw, Cline:2014dwa, Hoenig:2014dsa} where $l=e,\mu$. If the dark photon couples directly to $B+L+D$, in this $m_A$ range ${\rm BR}(A\to l^+l^-)\approx 20\%$ and LHC measurements can exclude a fraction of the parameter space between $m_A=30$~GeV and 1 TeV, where the full allowed range for the coupling is $10^{-4} \lesssim g_A < 4\pi$, with the exception of a window around the $Z$ mass that is partially excluded by electroweak precision observables~\cite{Cline:2014dwa, Hoenig:2014dsa, Curtin:2014cca}.  The high luminosity LHC run at $\sqrt{s}=14$~TeV~\cite{Curtin:2014cca} has the potential to extend the reach on $g_A$ by almost a factor of 10 with DY searches. For masses up to a few TeV, and $\OO(1)$ couplings, searches for contact interactions can also play a role~\cite{Aad:2014wca, Khachatryan:2014cja}. However if we omit the direct coupling to the lepton current, most of the parameter space is unconstrained. In this case discovering the dark photon in this mass range might prove extremely hard since it would behave as a dijet resonance with a small coupling to the SM quarks.

Similarly, for 50 MeV $\lesssim m_A \lesssim$ 10 GeV, the main signals are expected in dileptonic final states at $B$-factories. The BaBar collaboration already performed a search for $e^+e^-\to \gamma A\to \gamma l^+l^-$~\cite{Lees:2014xha}. Reinterpretations of mono-photon signatures are also relevant~\cite{Essig:2013vha}. If we couple the dark photon to lepton number at tree-level, the BaBar search for leptonic final states drives the sensitivity in the whole mass range excluding $g_A \gtrsim {\rm few}\times10^{-4}$. However, as in the case of a heavy dark photon, we can evade this bound for $m_A \gtrsim 500$~MeV if we do not gauge lepton number. For $m_A \lesssim 500$~MeV even if the coupling to leptons only arises through kinetic mixing, the branching ratio to $l^+l^-$ is larger than $40\%$  when the decay to dark matter is not kinematically allowed. If it is, it typically dominates in the whole mass range under consideration, except in the vicinity of spin-one mesons~\cite{Tulin:2014tya} and constraints from leptonic final states vanish. Nonetheless monophoton searches become comparable in sensitivity in the range $3\;{\rm GeV}\lesssim m_A \lesssim 10$ GeV~\cite{Essig:2013vha}. Furthermore LSND~\cite{deNiverville:2011it} and the recently completed MiniBooNE run proposed in 2012~\cite{Dharmapalan:2012xp} become sensitive to a small fraction of our parameter space for $m_A\lesssim$ few hundred MeV. Note that the BaBar monophoton reinterpretation should be taken with a grain of salt given that missing energy was not a well behaved quantity as it is today at the LHC, due to the nature and calibration of the calorimeter. The reach of both BaBar searches can be extended considerably at Belle-II.  For example it was shown that a dedicated mono-photon trigger could improve the sensitivity of the dark matter search by approximately one order of magnitude~\cite{Essig:2013vha}. Finally, for MeV $\lesssim m_A \lesssim$ 50 MeV electron~\cite{Bjorken:2009mm, Bjorken:1988as, Riordan:1987aw, Bross:1989mp,Andreas:2012mt} and proton~\cite{Gninenko:2012eq, Bergsma:1985is, Essig:2010gu, Batell:2009di, Athanassopoulos:1997er,Blumlein:2011mv} beam dump experiments can exclude $g_A \gtrsim 10^{-8}$ and essentially all our parameter space, if visible decays dominate over dark matter decays. 

A large number of future dark photon searches~\cite{Essig:2010xa, Freytsis:2009bh, Wojtsekhowski:2009vz, Wojtsekhowski:2012zq, Beranek:2013yqa, Battaglieri:2014qoa, Izaguirre:2014bca, Kahn:2014sra} are also relevant for MeV $\lesssim m_A \lesssim $~GeV with typical sensitivities in the range $g_A\approx 10^{-3}-10^{-5}$. In particular HPS~\cite{Celentano:2014wya} has the potential to cover a considerable fraction of our parameter space between $m_A\approx$~few tens of MeV and $m_A\approx$~200 MeV if the decay to electrons dominates. We leave to future work the study of the complementarity of these searches and supernova cooling bounds that in our setup are modified by the possible dark photon decays to dark matter. In general bounds on the cooling of SN1987A~\cite{Raffelt:2006cw}, the Sun, horizontal branch stars and red giants~\cite{An:2013yfc}, of white dwarfs~\cite{Dreiner:2013tja} and neutron stars~\cite{Curtin:2014afa} can also place important constraints on our parameter space for eV~$\lesssim m_A \lesssim 100$~MeV. However we do not discuss astrophysical bounds in detail here.  We mention only that if estimated naively, constraints from SN1987A seem to exclude a fraction of the parameter space relevant for direct detection even when the dark photon mass is much larger than the supernova temperature.  Nonetheless once the mean free path of the dark matter is taken into account, we find that dark matter masses in the range interesting for direct detection ($m_{DM} \gtrsim 300$~MeV) are not affected, since the particles are trapped inside the supernova.

\section{Conclusion and Future Directions}

In this paper, we have presented a new mechanism through which an asymmetry in one sector of the theory generates an asymmetry in a different sector, as long as the two are connected by a current-current operator.  We have applied it to asymmetric dark matter theories showing that an asymmetry in the dark sector can be used to generate the observed baryon energy density. This approach is unique among models of asymmetric dark matter, since it does not require thermal contact between the two sectors and allows the dark matter mass to become both extremely light or extremely heavy depending on whether the asymmetry is first generated in the dark sector or in the visible sector. Since thermal contact with the SM is not required, we can easily have realizations of the mechanism where it is not necessary to annihilate away the symmetric component of dark matter that might never have been sizable during the history of the universe.  This approach also features an Asymmetric WIMP miracle where if the dark matter and gauge boson mass are around the weak scale, the SM baryon number abundance is explained. 

The mechanism allows for considerable model building freedom even within the boundaries of baryogenesis. We have shown that it can be realized both if the dark matter is described by a set of particles or if it is described by a classical field. Furthermore the details of how $B$ or $L$ violation is included in the model and the temperature at which SM and dark sector decouple are not crucial. In practice this can lead to a variety of explicit realizations with potentially rich phenomenologies. Here we have discussed only a particular realization involving particle dark matter, where the SM and dark sector are in thermal equilibrium until $T\approx 100$~GeV and $B+L$ violation is provided entirely by the SM electroweak sphalerons. We found that the gauge boson connecting SM and dark sector could be detected at the LHC, $B$-factories or future beam dump experiments and we expect direct detection signals that are in the range of sensitivity of SuperCDMS SNOLAB for $m_{DM} > 300$~MeV.  This leaves open the exploration of models where the dark matter is a classical asymmetric field, where different sources of $B$ and $L$ violation are included or where the current-current interaction is not generated by the tree-level exchange of a vector boson.

\section*{Acknowledgments}
We thank Kfir Blum, Tim Cohen, Mariangiela Lisanti, Nima Arkani-Hamed and Josh Ruderman for carefully reading the manuscript and for useful comments. We also thank Philip Schuster and Natalia Toro for discussions that led to this project. RTD is supported by the NSF grant PHY-0907744.  AH is supported by the DOE grant DE-SC0009988.

\appendix

\section{A baryon asymmetry from chemical potentials}\label{sec:asym}

In the presence of a chemical potential, thermal equilibrium carries a net charge given by
\bea
n_{b}-n_{\bar b} &=& \sum_i q_i g_i \int \frac{d^3p}{(2\pi)^3} \left( \frac{1}{e^{(p-\mu_i)/T} + 1} - \frac{1}{e^{(p+\mu_i)/T} + 1} \right) \nn \\
&\approx& \sum_i q_i g_i \left( \frac{\mu_i^3}{6 \pi^2}  + \frac{\mu_i T^2}{6} \right) \approx \sum_i q_i g_i  \frac{\mu_i T^2}{6}, \quad T\gg m_i , \mu_i \, . \label{eq:ab}
\eea
where $T$ is the temperature, $q_i$ the baryon charge of the particle, $m_i$ its mass, $\mu_i$ its chemical potential, $g_i$  the number of degrees of freedom a particle has, and the sum goes over particles but not anti-particles. In our model we have both an explicit term in the Lagrangian that acts as a chemical potential and the familiar thermal chemical potentials of an equilibrium thermodynamical system. In this appendix we compute the baryon asymmetry generated in the SM in the presence of this extra explicit chemical potential.

As mentioned in the introduction, electroweak sphalerons violate $B+L$.  Therefore there are only two possibilities for the generation of the asymmetry, either only $B+L$ is violated or both $B+L$ and $B-L$ are.  In the first case, the $B+L$ violating sphalerons are in equilibrium until $T_S \approx 132$~GeV~\cite{D'Onofrio:2014kta}.  On the other hand $B-L$ violation typically decouples at a much higher, model dependent temperature and if present would dominate the asymmetry generation.  In the following subsections, we discuss the two different possibilities in more detail. 

\subsection{$B+L$ violation only}

There are two distinct components of the chemical potential relevant to our discussion.  The first is the familiar chemical potential introduced if a thermal system carries a charge $Q$.  We call it the ``thermal" chemical potential.  The second component is due to an explicit term in the Lagrangian that goes as $\mu Q$.  We call it the ``explicit" chemical potential. One can view the difference between the thermal and explicit chemical potentials as follows.  In the zero temperature vacuum, there is no thermal chemical potential and only the explicit chemical potential is present.  The explicit chemical potential is a CPT violating interaction that splits the energy levels of particles and anti-particles.  We can interpret the Hamiltonian in a simple way by ignoring the connection between particle and anti-particle and consider an equivalent system with two particles with different masses and opposite charge\footnote{Technically, only the lowest energy levels of these two systems would be equivalent as chemical potentials and masses shift energy levels in different ways.}. When putting the system at finite temperature in the early universe, we introduce the thermal chemical potential.  As is familiar, any reaction that is in equilibrium has the sum of the thermal chemical potentials on one side equal to the sum on the other side.  Much like in the case of mass terms, the presence of the explicit chemical potential does not change these relations.
  
The explicit chemical potential appears in the Lagrangian via the operator in Eq.~(\ref{eq:JJ})
\bea
\LL \supset  \mu_B \psi_B^\dagger \psi_B + \mu_L \psi_L^\dagger \psi_L
\eea
where $\mu_B = (g_+ + g_-) n_{DM}/3\Lambda^2$ is the explicit chemical potential for the baryons and $\mu_L = (g_+ - g_-) n_{DM}/\Lambda^2$ for the leptons. Keeping $\mu_B \neq \mu_L$ provides a simple check of the final result.  If $B-L$ is a good symmetry, a chemical potential coupled to the $B-L$ current should have no effect.  Thus the baryon and lepton asymmetries must be proportional to $3 \mu_B + \mu_L$. To simplify matters we work in the limit $T \gg m, \mu$, where $m$ is the mass of any particle carrying baryon or lepton number and $\mu$ stands for both the explicit and thermal chemical potential. Using this approximation and Eq.~(\ref{eq:ab}), we can write the baryon and lepton asymmetries as
\bea
Y_B &\equiv& \frac{n_{b}-n_{\bar b}}{s} = \left( 3 (\mu_{uL} + \mu_{uR} + 2 \mu_B) + 3 (\mu_{dL} + \mu_{dR} + 2 \mu_B) \right) \frac{T^2}{6 s}\, , \nn \\ 
Y_L &\equiv& \frac{n_{l}-n_{\bar l}}{s} =  \left( 3 (\mu_{i} + \mu_{iL} + \mu_{iR} + 3 \mu_L) \right) \frac{T^2}{6 s}\, , \label{eq:baryoncharge}
\eea
where $s= 2 \pi^2 g_{\star S}(T) T^3/45$ is the entropy density, $\mu_{u(d)L}$ are the thermal chemical potentials of the left-handed up (down) type quarks, $\mu_{u(d)R}$ for the right-handed up (down) type quarks, $\mu_{iL(R)}$ for the charged leptons and $\mu_i$ for the left-handed neutrinos.  We have assumed that flavor mixing maintains the equality of the chemical potentials for the three generations of quarks. Following Ref.~\cite{Harvey:1990qw}, we denote the remaining Standard Model thermal chemical potentials as
\begin{itemize}
\item $\mu_W$ : $W^-$ gauge boson
\item $\mu_0$ : Neutral Higgs field
\item $\mu_-$ : Charged Higgs fields
\end{itemize}
The fact that electroweak processes are in equilibrium dictate the following relations between the Standard Model chemical potentials
\bea
\nonumber 
\mu_W = \mu_- + \mu_0 &\qquad& W^- \leftrightarrow H^- + H^0 \\
\nonumber 
\mu_{dL} = \mu_{uL} + \mu_W &\qquad& W^- \leftrightarrow \bar u_L + d_L \\
\nonumber 
\mu_{iL} = \mu_i + \mu_W &\qquad& W^- \leftrightarrow \bar \nu_{iL} + e_{iL} \\
\nonumber 
\mu_{uR} = \mu_0 + \mu_{uL} &\qquad& H^0 \leftrightarrow \bar u_L + u_R \\
\nonumber 
\mu_{dR} = -\mu_0 + \mu_W + \mu_{uL} &\qquad& H^0 \leftrightarrow  d_L + \bar d_R \\
\mu_{iR} = -\mu_0 + \mu_W + \mu_i &\qquad& H^0 \leftrightarrow e_{iL} + \bar e_{iR} \label{eq:chempot}
\eea
Furthermore we introduce $B+L$ violation by requiring
\be
3 ( \mu_{uL} + 2 \mu_{dL} ) + 3 \mu_i = 0\, .
\ee
This relation is usually attributed to electroweak sphalerons, but chemical potentials are insensitive to the details of $B+L$ violation. Adding to the theory any other $B+L$ violating interaction would give the same result.

After $B+L$ violation freezes out, baryon and lepton number are good symmetries of the Standard Model and any existing asymmetry cannot be changed. What determines the abundances observed today are the baryon and lepton number abundances when $B+L$ violation freezes out. 
If this happens after the electroweak phase transition, in addition to the relations in Eq.~(\ref{eq:chempot}), we set $\mu_0 = 0$. $\mu_0$ is the thermal chemical potential of the neutral component of the Higgs doublet that becomes zero due to the vacuum condensate of $H^0$. We also set the total electric charge
\be
Y_Q &=& \left( 6 (\mu_{uL} + \mu_{uR} + 2 \mu_B) - 3 (\mu_{dL} + \mu_{dR} + 2 \mu_B) - 3 (\mu_{iL} + \mu_{iR} + 2 \mu_L ) -4 \mu_W -2 \mu_- \right) \frac{T^2}{6 s}\, , \nn \\ 
\ee
of the universe to zero because inflation dilutes away any initial charge. Solving the equations in~(\ref{eq:chempot}) for the thermal chemical potentials and using these two additional constraints ($\mu_0=0$ and $Y_Q=0$), we can rewrite Eq.~(\ref{eq:baryoncharge}) as
\bea \label{eq:Tsmall}
Y_B &=& \frac{6}{37} \left(2 Y_{B-L} + 17 (3 \mu_B + \mu_L)  \frac{T^2}{6 s} \right)\, , \nn \\ 
Y_L &=& \frac{1}{37} \left(-25 Y_{B-L} + 102 (3 \mu_B + \mu_L)  \frac{T^2}{6 s} \right)\, .
\eea
Here we allow the presence of an initial $B-L$ asymmetry in view of the results in the next subsection, where we introduce $B-L$ violation at higher temperatures.
If we assume that $T_{S}$, the temperature at which electroweak sphalerons freeze-out, is smaller than the temperature at which the electroweak symmetry is broken, $T_{EW}$, as suggested by current data~\cite{D'Onofrio:2014kta}, the results in Eq.~(\ref{eq:Tsmall}) are the only relevant ones in the absence of $B-L$ violation. 

This result neglects corrections proportional to $m_t/T$. To incorporate them we can write the top quark contribution to the baryon asymmetry as
\be
\frac{\left(n_b - n_{\bar b}\right)_{\rm top}}{s}\equiv f\left(\mu, T, m_t\right)\approx \frac{\mu}{T}f_0\left(T, m_t\right)\, ,
\ee
where we have assumed that $T\gg \mu$ still holds, but allowed for $m_t\sim T$. In this limit we still have a system of linear equations for the thermal chemical potentials. The new equations can be easily obtained from the ones in~(\ref{eq:chempot}) by changing the coefficients of the top quark contribution. If we define 
\be
\delta_t \equiv \frac{f_0\left(T, m_t\right)}{f_0\left(T, 0\right)} =\frac{\int_{m_t}^\infty dE \frac{E \sqrt{E^2-m_t^2}}{\cosh^2(E/2T)}}{\int_0^\infty dE \frac{E^2}{\cosh^2(E/2T)}}=\frac{3}{2\pi^2 T^3} \int_{m_t}^\infty dE \frac{E \sqrt{E^2-m_t^2}}{\cosh^2(E/2T)}
\ee
then the results in~(\ref{eq:Tsmall}) are generalized to
\bea
Y_B &=& 6\frac{(5 + \delta_t) Y_{B-L} + (38 + 13 \delta_t) (3 \mu_B + \mu_L) \frac{T^2}{6 s}}{97 + 14 \delta_t}\, , \nn \\ 
Y_L &=& \frac{-(67 + 8 \delta_t) Y_{B-L} +  6 (38 + 13 \delta_t) (3 \mu_B + \mu_L) \frac{T^2}{6 s}}{97 + 14 \delta_t}\, . \label{eq:Tsmallmt}
\eea
For $\delta_t=1$ we recover the previous equations. In the example discussed in Section~\ref{sec:sphalerons} where electroweak sphalerons generate the asymmetry, $\delta_t \approx 0.8$.

\subsection{$B-L$ violation}\label{sec:BmL}
In this subsection, we assume that there is explicit $B-L$ violation at high energies. As stated above, the thermal chemical potentials are sensitive only to which symmetries are broken or preserved.  As a result, how $B-L$ is broken does not influence the equilibrium values of the chemical potentials. In the context of the Standard Model, the lowest dimensional operator that breaks $B-L$ is the familiar lepton number violating dimension five interaction that can generate neutrino masses:
\bea
\frac{(H l)^2}{\Lambda_{B-L}} \label{eq:dim5Lepton}\, . \label{eq:dim5BmL}
\eea
All dimension six operators compatible with the SM symmetries preserve $B-L$ and to break it through baryon number violation requires introducing dimension eight operators. Adding to the SM Lagrangian the operator shown in Eq.~(\ref{eq:dim5BmL}) allows the reaction
\bea
H^- + e_{L}^+ \leftrightarrow H^+ + e_{L}^-
\eea
which sets $\mu_- = \mu_{iL}$, where $\mu_-$ is the thermal chemical potential of $H^-$.  All the relations between chemical potentials listed in~(\ref{eq:chempot}) and used in the previous section remain valid. Here we consider $B-L$ violating interactions that freeze-out above $T_{EW}$. For example, for the operator in Eq.~(\ref{eq:dim5Lepton}), $m_\nu \lesssim 0.1$~eV implies $\Lambda_{B-L} \gtrsim 10^{14}$~GeV which gives a freeze out temperature of $T_{B-L} \sim \Lambda_{B-L}^2/M_{Pl}\approx 10^{10}$~GeV.  When $T > T_{EW}$ we can set both the electromagnetic charge and the weak isospin charge, 
\be
Y_{Q_3} &=& \left( \frac{9}{2} (\mu_{uL} - \mu_{dL} ) + \frac{3}{2} ( \mu_i - \mu_L ) - 4 \mu_W - (\mu_0 + \mu_-) \right) \frac{T^2}{6 s}\, ,
\ee
to zero since $SU(2)_L$ is a good symmetry of the theory and inflation diluted away initial charges.  Since at $T > T_{EW}$, the Higgs field has not obtained a vev, $\mu_0$ can be non-zero. Repeating the previous calculation with these new constraints, we obtain
\be
Y_{B-L} = \frac{3}{11} ( 28 \mu_B - 17 \mu_L )  \frac{T^2}{6 s}\, . \label{eq:BmL}
\ee
After the electroweak phase transition, this can be translated into a baryon number abundance via Eqs.~(\ref{eq:Tsmall}) or (\ref{eq:Tsmallmt}). Since $T_{B-L}$ is much larger than $T_{S}$, we can neglect the sphaleron contribution to $Y_B$ and keep only the term proportional to $Y_{B-L}$ in Eq.~(\ref{eq:Tsmall}). 

\bibliographystyle{utphys}
\bibliography{ref}

\end{document}